\documentclass[doublecol]{epl2} 

\newcommand{\beq}{\begin{equation}}
\newcommand{\eeq}{\end{equation}}
\newcommand{\bea}{\begin{eqnarray}}
\newcommand{\eea}{\end{eqnarray}}
\def\eqref#1{(\ref{#1})}

\begin{document}

\title{From quantum foam to graviton condensation:\\ the Zel'dovich route}

\author{
Pablo G. Tello\inst{1} \and Sauro Succi\inst{2,3,4} \and Donato Bini\inst{2,5}\and   Stuart Kauffman\inst{6} 
}

\institute{                    
\inst{1} CERN, Geneva, Switzerland\\
\inst{2}Istituto per le Applicazioni del Calcolo \lq\lq M.~Picone,\rq\rq CNR, I-00185 Rome, Italy\\
\inst{3}
Istituto Italiano di Tecnologia, 00161 Rome, Italy\\
\inst{4}Physics Department, Harvard University, Cambridge,USA\\
\inst{5}INFN, Sezione di Roma Tre, I-00146 Rome, Italy\\
\inst{6}Institute for Systems Biology, Seattle, WA 98109, USA\\
}
\abstract{
Based on a previous ansatz by Zel'dovich for the gravitational energy of virtual 
particle-antiparticle pairs, supplemented with the Holographic Principle, we
estimate the vacuum energy in a fairly reasonable agreement with the 
experimental values of the Cosmological Constant. 
We further highlight a connection between Wheeler's quantum foam
and graviton condensation, as contemplated in the quantum 
$N$-portrait paradigm, and show that such connection also leads to 
a satisfactory prediction of the value of the cosmological constant.
The above results suggest that the \lq\lq unnaturally" small value of 
the cosmological constant may find a quite \lq\lq natural" explanation once the nonlocal 
perspective of the large $N$-portrait gravitational condensation is endorsed. 
}

\maketitle

\section{Introduction}
In a previous Letter, these authors argued that the Zel'dovich 
picture of gravitational energy as due to the ceaseless 
generation and annihilation of 
particle-antiparticle pairs, once combined with
the holographic principle, provides a natural and quantitative 
account of the exceedingly small value of the
cosmological constant.  
More precisely, it was shown that the combination of Zel'dovich picture 
with the holographic principle predicts
a cosmological constant of the order of $\Lambda (l) l_{\rm P}^2 \sim (l_{\rm P}/l)^2$,  $l_{\rm P}$ being
the Planck length and $l$  the  infrared scale.  By taking $l$ of the order of the size of the universe,
this delivers $\Lambda l_{\rm P}^2 \sim 10^{-124}$ in close agreement with the observed value $10^{-122}$.

In this paper, we push the idea further ahead by showing that the Zel'dovich picture also
permits to draw a consistent connection between a revised version of Wheeler's quantum foam 
\cite{Misner:1973prb,Carlip:2022pyh}
and the large $N$-portrait framework, according to which the cosmological 
constant is associated with  the
Bose-condensation of a gas of ultrasoft gravitons permeating the entire Universe
\cite{Dvali:2011aa}. 

In passing, we note that the connection between Wheeler's quantum foam and the large N-portrait
via the Zel'dovich-holographic scenario, provides a very \lq\lq natural" explanation for the allegedly \lq\lq innatural"
smallness of the cosmological constant.   Such smallness is indeed due to the nonlocality of the
gravitational excitations which, although generated at the Planck scale,  do nonetheless persist
and extend across the entire size of the Universe.  
In this respect,  the alleged \lq\lq unnaturalness" of the cosmological constant appears as an 
artifact of local quantum field theories, as opposed to the nonlocal $N$-portrait framework,
along the lines envisaged in \cite{EPJSS}.

\section{Wheeler's quantum foam}

Back in 1955,  J.~A.~Wheeler argued that on account of the uncertainty principles, 
at the shortest scales of the order of the Planck length spacetime itself should fluctuate
as a sort of quantum foam \cite{Wheel55}.  Even though experimental evidence of
such quantum foam remains elusive, the notion of a fluctuating quantum spacetime has
attracted considerable interest ever since.
Wheeler's simplest claim for the possible existence of the quantum 
foam could be summarized as follows \cite{Misner:1973prb,Carlip:2022pyh}. 

Let us consider a cubic region of space of side $l$ and volume $l^3$. 
The smallest quantum gravitational excitation fitting within such a volume  
is a graviton of wavelength $l$, with energy  
$\hbar c/l$, implying an energy density given by
\beq
\label{eq:1}
\rho c^2 \sim \frac{\hbar c}{l^4 } \,.
\eeq
On the other hand, as it is known, from classical General Relativity, a small metric 
fluctuation, $\delta g$, within this cube features an effective energy density: 
\beq
\label{eq:2}
\rho c^2 \sim \frac{c^4}{G}\left(\frac{\delta g}{l}\right)^2,
\eeq
Equating Eqs. \eqref{eq:1}  and \eqref{eq:2}  and considering the definition of the Planck
 length, $l_{\rm P}^2=\hbar G/c^3$, leads to 
\beq
\label{eq:3}
\delta g (l) \sim \frac{l_{\rm P}}{l}\,,
\eeq
showing that the amplitude of metric fluctuations scales inversely with their size and become order one at the Planck scale.
Incidentally, we note that the inverse size dependency is a definite signature of singularity of the continuum
limit (Einstein's gravitation), since the gradient 
$\delta g(l)/l$  diverges like $1/l^2$ in the 
limit $l \to 0$.

During the past years, different models attempting to understand the emergence of the quantum foam have been proposed \cite{Hossenfelder:2012jw}. Arguably, two of them, among others, became significant candidates: the holographic model 
\cite{Karolyhazy:1966zz,Ng:1993jb,Ng:1995km,Ng:2003ag} and 
the random walk model \cite{Diosi:1989hy,Amelino-Camelia:1998mjq}. 

The holographic model became known as such given its consistency with the Holographic Principle \cite{tHooft:1993dmi,Susskind:1994vu}. Roughly summarized, it considers the operational resolution limits for the measurement of spacetime (e.g. length and/or time intervals), to derive the well-known expression 
$\delta l\ge  l^{1/3} l_{\rm P}^{2/3}$ for the length fluctuations. 

The random walk model considers the operational definition 
of the minimal distance, $l$, in a fuzzy spacetime affected by quantum fluctuations. 
These fluctuations are primarily characterized by their root-mean-square deviation 
$ \sigma_d = \langle(\delta l)^2\rangle^{1/2}$, the simplest proposal being $\sigma_d<l$. 
The model postulates the well-known expression 
$\delta l\ge  l^{1/2} l_{\rm P}^{1/2}$  for the length fluctuations. 
It is beyond the scope of this article to offer a detailed review of both models,
but an excellent account of them can be found in the aforementioned references.

Despite their phenomenological plausibility, both models are ruled out by 
recent experimental observations, which also set stringent 
limits to any phenomenological 
model considering the fuzziness of the spacetime 
at very small distances \cite{Perlman:2014cwa,Nemiroff:2011fk}.
A few  important remarks are in order, especially in relation to 
the holographic model.
The first point is that ruling the holographic model out does not 
necessarily imply the demise of the Holographic Principle \cite{Perlman:2014cwa}. 
Thus, a few questions remain: what is the origin of the 
metric fluctuations at the Planck scale? Can they be understood without 
invoking operational considerations related to the measurement of spacetime? 
Can the Holographic Principle still play a fundamental role in accounting for
such fluctuations? 

The second remark concerns the origin of dark energy and its implications for the small experimental value of the Cosmological Constant. The holographic model plausibly opened the possibility to relate holography and dark energy (holographic dark energy). Its demise thus,  prompts out additional questions. 
How could dark energy be explained in relation to the Holographic Principle without invoking the holographic model? And with it, how might it be related to the experimental value of the Cosmological Constant?

In this article,  we address the above questions by revisiting a previous 
ansatz due to Zel'dovich and connecting it with the Holographic Principle.

\section{The Vacuum Catastrophe}

The so-called Vacuum Catastrophe refers to the astronomical discrepancy 
between the observed energy density of the vacuum and the one 
predicted by quantum field theory. 
It has been termed as the \lq\lq worst prediction in physics.\rq\rq $\,$ The predicted density is $\sim 10^{120}$ vs. $\sim 10^{-9}\, J/m^3$ estimated by current measurements \cite{14,Bousso:2007gp,Weinberg:1988cp,Carroll:2000fy,Sola:2013gha,Klinkhamer:2008ns}.
Ignoring numerical factors, the theoretical vacuum energy density, $\rho_v$, is expressed as 
\beq
\label{eq:4}
\rho_v c^2 \sim  \hbar \omega  \frac{c^3}{\omega^3}\,,
\eeq
where $\omega$ is the radiation frequency of the vacuum modes 
and the other symbols are standard.
Let us assume, in connection with Wheeler, vacuum modes with a 
frequency $\omega\sim  c/l$ associated with gravitons.  
The expression \eqref{eq:9} rewrites as 
\beq
\label{eq:5}
\rho_v c^2 \sim \frac{\hbar c}{l^4}\,,
\eeq
which is identical to Eq. \eqref{eq:1}. 
Therefore, applying the same heuristic argument as Wheeler, the expression \eqref{eq:5} leads to the same metric fluctuations at the Planck scale given by \eqref{eq:3}. 
In the following section we discuss further how these metric fluctuations might 
be due to the gravitational energy of virtual particle-antiparticle pairs, 
continually generated and annihilated in the vacuum state.

\section{Zel'dovich's ansatz and the Holographic Principle}

Recently, the authors have revisited Zel'dovich's ansatz to  provide an explanation of the current value of the Cosmological Constant,  see \cite{Tello:2022dtz}.  
Here, a brief reminder is made for the sake of consistency. 
Zel'dovich argued that, since the bare zero-point energy is unobservable, the observable 
contribution to the vacuum energy density, $\rho_v c^2$, is given by the gravitational 
energy of virtual particle-antiparticle pairs ceaselessly 
generated and annihilated in the vacuum state \cite{20,Zeldovich:1968ehl}. Therefore, 
\beq
\label{eq:6}
\rho_vc^2 \sim \frac{ Gm^2 (l) }{l}\frac{1}{l^3}\,.
\eeq
In the expression above, also according to Zel'dovich, the vacuum contains excitations 
with an effective density $m(l)/l^3$. 
Additionally, by considering the Compton's expression for the wavelength, the 
effective mass of the particles at scale $l$ is taken as $m(l)\sim  \hbar /(cl)$. 
This leads to an energy density 
\beq
\label{eq:7}
\rho_v c^2\sim \frac{G\hbar^2}{c^2 l^6} \,.
\eeq
Equating Eqs. \eqref{eq:2} and \eqref{eq:7}, leads to 
\beq
\label{eq:8}
\delta g \sim \left(\frac{l_{\rm P}}{l}\right)^2\,.
\eeq

This shows a steepest (more singular) inverse size dependence than predicted by Wheeler,  
though still leading to metric fluctuations of order one at the Planck scale.

\subsection{Connection of the vacuum energy with the cosmological constant}

The connection of this ansatz with the Cosmological Constant goes as 
follows \cite{Carroll:2000fy}. Let us define a local Cosmological Constant as 
\beq
\label{eq:9}
\Lambda(l)\sim \frac{G}{c^2}  \rho_v (l) \,. 
\eeq
By considering Eqs. \eqref{eq:6} and \eqref{eq:7}, one readily obtains 
\beq
\label{eq:10}
\Lambda(l)l_{\rm P}^2\sim \left(\frac{l_{\rm P}}{l}\right)^6\,.
\eeq
Reasoning further, the steep $1/l^6$ dependence implies that $\Lambda(l)$ 
is largely dominated by  the chosen  UV cutoff, $l_{\rm UV}$.
This suggests to rewrite Eq. \eqref{eq:10}  as 
\beq
\label{eq:11}
\Lambda l_{\rm P}^2\sim \left(\frac{l_{\rm P}}{l_{\rm UV}} \right)^6\,.
\eeq
By considering the Holographic Principle in the form 
of $l_{\rm UV}=l^{1/3} l_{\rm P}^{2/3}$,  one obtains: 
\beq
\label{eq:12}
\Lambda l_{\rm P}^2\sim \left(\frac{l_{\rm P}}{l}\right)^2\,,  
\eeq
and further 
\beq
\label{eq:13}
\Lambda \sim \frac{1}{l^2} \,.
\eeq
Taking $l$ as the current radius of the Universe, gives a value of $\Lambda\sim 10^{-54}\,m^{-2}$, fairly comparable, given the approximate nature of the assumptions taken, to the experimental value $\Lambda \sim 10^{-52}\, m^{-2}$ \cite{WMAP:2003elm,WMAP:2006bqn}. An attempt to explain why this value provides such a remarkable approximation is made in the following section, also in relation with the questions formulated before. The expression obtained for the Cosmological Constant is put into relation with postulated theories involving graviton condensation.

\section{Connections with graviton condensation  theories}

Einstein's General Relativity (GR) is a classical theory of gravity. 
From the quantum point of view, it propagates a unique weakly coupled quantum particle with zero mass and spin-2. Recently, the so-called black hole quantum $N$-portrait paradigm, postulates a quantum self-coupling of gravitons consistently defined as \cite{Dvali:2011aa,Dvali:2012en,Dvali:2012rt,Dvali:2012gb,Dvali:2010bf,Casadio:2013hja,Dvali:2010jz}
\beq
\label{eq:14}
\alpha_g=\left(\frac{l_{\rm P}}{l}\right)^2 \,,
\eeq
where $l$ is a classical wavelength at low energies. 

The idea behind this approach is that Einstein gravity, viewed as a quantum field theory, becomes self-complete as
it prevents from probing distances shorter than the Planck length, $l_{\rm P}$,  
by producing a large occupation number $N$ of very long wavelength $l\gg  l_{\rm P}$ to any high-energy scattering. 

The merit behind the black hole's quantum $N$-portrait is that it offers an explanation 
of black hole properties, such as thermality and Bekenstein's entropy, which was believed 
to be impossible within the existing framework of Einsteinian 
gravity, no matter whether classical or quantum. 
The explanation of the Bekenstein's entropy also establishes a link with the Holographic Principle. 
In the context of this article, it is worth noticing that the expression \eqref{eq:12} directly 
determines the postulated quantum self-coupling of gravitons and its relationship to the Cosmological Constant:
\beq
\label{eq:15}
\Lambda \sim \frac{1}{l_{\rm P}^2} \left(\frac{l}{l_{\rm P}} \right)^2\sim \frac{\alpha_g}{l_{\rm P}^2}\,, 
\eeq
which, multiplying both sides by $l_{\rm P}^2$, becomes 
exactly Eq. \eqref{eq:14}, thereby delivering 
the postulated quantum self-coupling of gravitons:
\beq
\label{eq:16}
\Lambda l_{\rm P}^2\sim \left(\frac{l_{\rm P}}{l}\right)^2\sim \alpha_g\,.
\eeq

The above expression portrays the gravitons as the \lq\lq sinews" of gravity,  i.e.,   coherent structures
transversally confined within a squarelet of area $l_{\rm P}^2$ on the surface of 
a sphere of radius $l$, and fully delocalized  along the longitudinal direction, basically 
connecting different points on the sphere along its diameter of size $l$ (nonlocality).

Consistently with this picture, the  graviton counting is then 
$N l_{\rm P}^2 l \sim l^3$,  namely $N \sim (l/l_{\rm P})^2$.    

So far, we have shown that the Vacuum Catastrophe leads to a similar expression for 
the metric fluctuations at the Planck scale as those postulated by Wheeler, under the assumption 
of vacuum modes associated with gravitons.

These fluctuations have been linked to gravitational interactions between virtual particle-antiparticle 
pairs through the proposed connections between Zel'dovich's ansatz and the Holographic Principle,  without 
invoking any operational limitations associated to the measurement of spacetime.  
Furthermore, the approach taken so far seems to reasonably agree with the experimental values of the 
Cosmological Constant and links up naturally to the quantum 
self-coupling of gravitons postulated by the quantum $N$-portrait hypothesis. 

A question remains, though, as to why one should choose $l$ comparable to the Universe radius.
As explained elsewhere \cite{Medved:2008vh}, considering the possibility for dark energy to 
be associated with \lq\lq quanta of gravity\rq\rq, plausibly 
implies that these quanta should be
naturally delocalized across the full extent of the Hubble sphere, 
which in our case is the Universe radius.
An interesting side remark concerns the large occupation number $N$ mentioned above.  
According to the quantum $N$-portrait hypothesis, $N$ is given by 
\beq
\label{eq:17}
N=\frac{1}{\alpha_g} \,, 
\eeq
By taking again $l_{\rm P}\sim 10^{-35}\, m$ and $l\sim 10^{27}\, m$ gives $N\sim 10^{120}$, corresponding 
to a very large occupation number $N$ of a very long wavelength compared to the Planck one and
coinciding with the order of magnitude difference between the measured value of 
the Cosmological Constant and the one obtained theoretically.

Another relevant remark is as follows.  As mentioned before, while gravitons are assumed
to be massless, they can still carry energy, like other gauge quanta, such as photons and gluons, do. 
The relation of the possible mass of a graviton, $m_G$, and the Cosmological Constant 
has been postulated elsewhere as 
\cite{Ali:2014qla,Das:2014sia,Das:2014agf,Das:2015dca}:
\beq
\label{eq:16bis}
m_G\sim \frac{\hbar}{c}\Lambda^{1/2}\,.
\eeq
Considering the expression for $\Lambda$ obtained previously in \eqref{eq:16},  
we obtain:
\beq
\label{eq:17bis}
m_G\sim \frac{\hbar}{cl }\,,
\eeq   
which again, by taking $l$ as the current radius of the Universe, gives a value of $m_G \sim 10^{-69}\, Kg$, 
in agreement with estimates obtained elsewhere of $m \sim 10^{-68}\, Kg$ 
(or $10^{-32}\, eV$) \cite{Ali:2014qla,Das:2014sia,Das:2014agf,Das:2015dca}. 

Expression \eqref{eq:17bis} has been postulated in the context of a modified nonlocal theory of gravity, 
specifically, the quantum corrected Raychaudhuri equation, in which geodesics are replaced 
by quantum Bohmian trajectories. 
Our estimate does not need any such modifications, nor does it assume any
non-locality through Bohmian trajectories. Instead, it naturally aligns with the proposed 
quantum $N$-portrait paradigm, in which \lq\lq quanta of gravity\rq\rq are naturally 
delocalized across the entire Hubble distance \cite{Medved:2008vh}.

\section{Towards Dark Energy as a Bose-Einstein Graviton Condensate}

As postulated by the quantum $N$-portrait, the strength of graviton-graviton interaction 
is measured by a dimensionless coupling constant given by expression \eqref{eq:14}, which 
could be interpreted as the relativistic generalization of the Newtonian attraction  potential 
among two gravitons, written as \cite{Dvali:2012en}:
\beq
\label{eq:18} 
V(l)=\alpha_g \; \frac{\hbar c}{l}\,.
\eeq
Multiplying expression \eqref{eq:6} by $l^3$, directly gives
\beq
\label{eq:19}
V(l)=\rho_v c^2 l^3 \sim \frac{G\hbar^2}{c^2 l^3}
= \hbar c \left(\frac{l_{\rm P}}{l}\right)^2\frac{1}{l}
= \alpha_g \; \frac{\hbar c}{l}\,,
\eeq
which is exactly expression \eqref{eq:18}. It is noticeable that, while expression \eqref{eq:18} has 
been estimated for the case of black holes, a similar one is obtained in the context of this paper 
in relation to dark energy and therefore the cosmological constant. 

In passing, we note that this also reminds electrostatic interactions between
relativistic electrons in graphene, which write $V_{el}(l) = \frac{e^2}{l} = \alpha_f \frac{\hbar c}{l}$,
where $\alpha_f = \frac{e^2}{\hbar v_f}$ is the graphene fine-structure constant, $v_f \sim c/100$
being the Fermi speed of the electron excitations.
Indeed, holographic analogies have been invoked in the context of electronic transport
in graphene \cite{HoloGraph}, which, due to the strong electronic coupling, $\alpha_f \sim 1$, can 
often be treated by classical hydrodynamic analogies \cite{Mendoza}.   

An equally interesting similarity with the quantum $N$-portrait concerns the large occupation 
number $N$, which indicates the extent to which quantum states are filled up by the excitations
of a quantum-mechanical system consisting of many identical gravitons sharing the same quantum state. 

According to the quantum $N$-portrait hypothesis, $N$, is given by 
\beq
\label{eq:17n}
N=\frac{1}{\alpha_g} \,,  
\eeq
By taking again $l_{\rm P}\sim 10^{-35}\, m$ and $l\sim 10^{27}\, m$ gives $N\sim 10^{120}$, corresponding 
to a large occupation number $N$ of very long wavelength 
compared to the Planck one. 
As interestingly noted elsewhere, this shows that the contribution of the vacuum energy density is 
strongly suppressed by large value of $N$ \cite{Kuhnel:2014gja}. 

This could explain the small value of the Cosmological Constant, as well as the reasonable agreement 
we obtained with its experimental value, since the suppression of $\Lambda$ 
by the number of gravitons would be of the order $1/N\sim 10^{-120}$. 

The above considerations seem to indicate that, in line with 
the quantum $N$-portrait, the Zel'dovich scenario also
involves graviton condensates with large occupation numbers.  
As this number increases, fueled by the ceaseless production of particle-antiparticle
pairs in the quantum foam, collective effects become dominant to the point
of triggering coalescence and self-condensation.
Such scenario provides a physical realization of the inverse cascade 
envisaged in \cite{EPJSS} on purely speculative grounds.  

It is worth referring to G. Dvali and C. Gomez for an explanation \cite{Dvali:2012en}. 
As these authors point out, it is useful to imagine localizing as many gravitons as 
possible within a space region of size $l$, in our case the Universe radius. 
In other words, trying to form a Bose-Einstein graviton condensate of characteristic wavelength $l$ 
by gradually increasing their occupation number $N$. 
When $N$ is small, the graviton interaction is negligible, but  
as $N$ increases, individual gravitons feel a stronger and stronger 
binding potential and at a critical occupation number, a self-sustained
condensate forms. 
The quantum $N$-portrait predicts a critical occupation number given precisely 
by $N_c=1/\alpha_g$,  which in our case leads to
\beq
\label{eq:18n}
N_c=\frac{1}{\Lambda l_{\rm P}^2} \sim \left(\frac{l}{l_{\rm P}} \right)^2 \,.
\eeq
The critical occupation number also indicates that the graviton 
condensate is maximally packed. 
Following the quantum $N$-portrait paradigm, the spectrum of fluctuations is 
determined by the Bogoliubov-De Gennes equation. 
The energy gap to the first Bogoliubov level, $\epsilon_1$, is then given by 
\beq
\epsilon_1=\frac{1}{N^{1/2}} \frac{\hbar c}{l}\,,
\eeq
which, for the case of the present paper, delivers  
\beq
\label{eq:20}
\epsilon_1=\hbar c \frac{l_{\rm P}}{l^2}\,.
\eeq
Considering that the Planck energy is $E_{\rm P}=\hbar/t_{\rm P}$, being $t_{\rm P}$ the Planck 
time, expression \eqref{eq:20} leads to:
\beq
\label{eq:21}
\epsilon_1=E_{\rm P} \left(\frac{l_{\rm P}}{l}\right)^2 \,.
\eeq
It is also interesting to notice that expression \eqref{eq:18n} resembles the 
Bekenstein bound, thereby connecting with the Holographic Principle and consequently 
to a holographic nature of dark energy \cite{Wang:2016och}.

\section{Conclusions}

Since hypothesized by Wheeler, quantum foam has generated a rich stream
of physics speculations related to the nature of fluctuations of 
spacetime at the fundamental level. 
Recent experimental observations, though,  indicate that previous models, based on 
plausible assumptions on the operational definition of length fluctuations 
at the Planck scale, should nevertheless be ruled out. 
The present paper offers an alternative angle to explain the origin of the so-called quantum 
foam and its potential relation to dark energy and 
the resulting experimental value of the Cosmological Constant.
 
The starting point in section 2, has been the assumption of gravitons with frequency 
by $\omega \sim c/l$ which lead to quantum foam metric fluctuations, as suggested by Wheeler 
when directly inserted into the expression of the vacuum energy density. 
In section 3, following Zel'dovich's ansatz, supplemented with the Holographic Principle, we have argued 
that such postulated gravitons are the mediators of the gravitational interaction between 
virtual particle-antiparticle pairs, continually generated and annihilated in the vacuum state. 
Furthermore,  by considering the vacuum energy density, a satisfactory estimate of the 
experimental values of the Cosmological Constant has been obtained, upon assuming $l$ as the Universe radius. 

In section 4, the assumption of such gravitons has been cross-checked with 
the quantum $N$-portrait paradigm, which predicts a graviton coupling for the case of black holes. 
It has been argued that the predicted coupling is also plausibly applicable to our case, related 
to the Cosmological Constant estimate obtained in section 3.

In section 5, our approach has been cross-checked against the mass of gravitons 
postulated by a modified theory of gravity based on a quantum corrected Raychaudhuri equation, in which 
geodesics are replaced by nonlocal Bohmian trajectories.  
We have argued that the postulated mass could be obtained by 
following the Zel'dovich approach with no need to invoke any 
gravity modifications. 
Moreover, in such an approach, non-locality emerges in the form
of \lq\lq quanta of gravity\rq\rq  delocalized across the entire 
Hubble distance, thus accounting for the match with the 
experimental value of the Cosmological Constant when 
taking $l$ as the Universe radius.

Finally, we have suggested that the approach discussed in this paper
is conducive to a Bose-Einstein graviton gas condensate, characterized 
by very large occupation numbers, which would 
explain the small value of the Cosmological Constant as due to the
suppression by the large  number of gravitons, 
of the order $1/N\sim 10^{-120}$. 

In summary, it seems plausible to consider the Zel'dovich's ansatz 
supplemented with the Holographic Principle,  to explain
the origin of quantum foam, dark energy and the small experimental 
value of the Cosmological Constant. 

It remains of course for experimental observations to validate or 
disprove the picture presented here. 
At the moment, its main merit is simplicity,  as it only 
necessitates two basic assumptions.
First, as proposed by Zel'dovich, that quantum vacuum fluctuations are linked 
to gravitational interactions between virtual particle-antiparticle pairs.  
Second, the existence of a UV cut-off dictated by the Holographic Principle.

\acknowledgments
DB acknowledges sponsorship of the Italian Gruppo Nazionale 
per la Fisica Matematica (GNFM) of the Istituto Nazionale di Alta Matematica (INDAM).
All authors are grateful to Prof. David Spergel for very valuable remarks and criticism.

\end{document}